\documentclass[fleqn,usenatbib]{mnras}

\usepackage{newtxtext,newtxmath}

\usepackage[T1]{fontenc}

\DeclareRobustCommand{\VAN}[3]{#2}
\let\VANthebibliography\thebibliography
\def\thebibliography{\DeclareRobustCommand{\VAN}[3]{##3}\VANthebibliography}

\usepackage{bm}
\usepackage{multirow} 
\usepackage{longtable} 
\usepackage{rotating}
\usepackage{multicol}
\usepackage{subfigure}
\usepackage{booktabs}
\usepackage{graphicx}	
\usepackage{amsmath}	

\usepackage{amssymb}	
\usepackage{orcidlink}  

\usepackage{xargs}
\usepackage{float}
\usepackage[pdftex,dvipsnames]{}

\usepackage{ulem}

\newcommand{\st} {\color{black} \sout}

\newcommandx{\unsure}[2][1=]{\todo[linecolor=red,backgroundcolor=red!25,bordercolor=red,#1]{#2}}
\newcommandx{\change}[2][1=]{\todo[linecolor=blue,backgroundcolor=blue!25,bordercolor=blue,#1]{#2}}
\newcommandx{\info}[2][1=]{\todo[linecolor=OliveGreen,backgroundcolor=OliveGreen!25,bordercolor=OliveGreen,#1]{#2}}
\newcommandx{\improvement}[2][1=]{\todo[linecolor=Plum,backgroundcolor=Plum!25,bordercolor=Plum,#1]{#2}}
\newcommandx{\thiswillnotshow}[2][1=]{\todo[disable,#1]{#2}}

\newcommand{\kms}{\mbox{\,km s}^{-1}}

\newcommand{\nth}{{n_{{\rm th}}}}

\newcommand{\dif}{\mathrm{d}}
\newcommand{\eff}{\epsilon_{\rm ff}}

\newcommand{\tauff}{{\tau_{\rm ff}}}

\title[The efficiency per free-fall time in collapsing cores]{The efficiency per free-fall time as a ratio of the Star Formation Rate to the gas-infall rate in collapsing cores: dependence on the core definition, accretion, and radial structure}

\author[Quesada-Zúñiga, F. et al.]{Fabián Quesada-Zúñiga\orcidlink{0009-0000-3167-2713},$^{1}$
Manuel Zamora-Avilés\orcidlink{0000-0002-2133-9973},$^{1,2}$
Enrique Vázquez-Semadeni\orcidlink{0000-0002-1424-3543},$^{3}$\thanks{E-mail: e.vazquez@irya.unam.mx}
Gilberto C. Gómez\orcidlink{0000-0003-4714-0636},$^{3}$
\newauthor
Aina Palau\orcidlink{0000-0002-9569-9234},$^{3}$ and Javier Ballesteros-Paredes\orcidlink{0000-0002-1081-9445}$^{3}$
\\
$^{1}$Instituto Nacional de Astrofísica, Óptica y Electrónica, Luis E. Erro 1, 72840 Tonantzintla, Puebla, México\\
$^{2}$Secretar\'{\i}a de Ciencia, Humanidades, Tecnolog\'{\i}a e Innovaci\'on (SECIHTI), Av. Insurgentes Sur 1582, 03940, Ciudad de M\'exico, M\'exico \\
$^{3}$Universidad Nacional Aut\'onoma de M\'exico, Instituto de Radioastronom\'ia y Astrof\'isica, Antigua Carretera a P\'atzcuaro 8701, Ex-Hda. San Jos\'e de la Huerta,\\ 58089, Morelia, Michoac\'an, M\'exico
}

\date{Accepted XXX. Received YYY; in original form ZZZ}

\pubyear{2026}

\begin{document}
\label{firstpage}
\pagerange{\pageref{firstpage}--\pageref{lastpage}}
\maketitle

\begin{abstract}

A parameter commonly used to characterise star formation activity in molecular clouds is the  efficiency per free-fall time, $\epsilon_{\rm ff}$, although commonly referred to as an efficiency, it is formally the ratio between the star formation rate (SFR) and the gas--infall rate. Here we numerically study the collapse of cores and define $\epsilon_{\rm ff} \equiv \langle \dot{M}_\star \rangle / (M_{\rm core} / \tau_{\rm ff})$, where $\langle \dot{M}_\star \rangle$ is the average  SFR, $M_{\rm core}$ is the gas mass within the core defined by a chosen density threshold, and $\tau_{\rm ff}$ is the free-fall time of the core gas. We perform simplified numerical experiments of the gravitational collapse of an isolated core, varying the initial mean number density ($ n_0 = 100 $ and $ 1000~\mathrm{cm^{-3}} $) and adopting open/closed boundary conditions to allow/disallow fresh gas accretion into the computational domain. We define the core as the gas cells above a density threshold. The simulations start with a slight central Gaussian overdensity that evolved into a power-law profile, $ n \propto r^{-p} $ with $ p \to 2 $. As the collapse proceeds, a sink particle forms in the centre of the core. We find that both the boundary conditions and the adopted core definition modify the measured core properties and, consequently, the inferred $ \epsilon_{\rm ff} $. Low-density models have less mass available in the numerical box, and their accretion histories are therefore much more sensitive to the choice of boundary conditions, while high-density runs, with their larger mass reservoirs, maintain similar accretion histories regardless of the boundaries. In all models, after sink formation, $\eff$ rises and then remains relatively stable while accretion continues to replenish the core's mass, but increases monotonically once the gas reservoir is exhausted. Somewhat counterintuitively, $\epsilon_{\rm ff}$ is systematically higher in the low-mass cores, since the larger gas infall rates onto the high-mass cores compensate for their higher star formation rates. We conclude that the inferred $\epsilon_{\rm ff}$ depends sensitively on both the adopted core definition and the external mass supply.

\end{abstract}




\section{Introduction}

 Dense cores embedded in molecular clouds (MCs) are the places where the star formation process takes place in our galaxy, as the final step of gravitational contraction ongoing in the clouds that leads to star formation \citep[e.g.,][]{Myers86}. Understanding the dynamical processes involved in the formation of dense cores is thus essential to understanding the mechanisms that drive star formation.

Regions known as ``clumps'' are defined as having average number densities of $n \sim 10^3 - 10^4$ cm$^{-3}$ and sizes of $\sim 0.1 - 2$ pc, while those known as ``cores'' have number densities of $n \geq 10^4$ cm$^{-3}$ and sizes of $\leq 0.1$ pc \citep[ e.g.,][and references therein]{chevance2022life}. Cores typically display radial power-law density profiles of the form $n \propto r^{-p}$, where  $p \sim 1.7 - 1.9$, with some deviations \citep[see an observational compilation in][]{Gomez_2021}. 

 Observational definitions of these structures rely on molecular-line and dust-continuum observations. Because molecular clouds are composed primarily of cold H$_2$, which is difficult to observe directly, their gas distribution is commonly inferred from emission by molecular tracers such as CO, NH$_3$, and N$_2$H$^+$, as well as from dust continuum emission \citep[e.g.,][]{Wong_2011, SanchezMonge2013_VLA-NH3-cores}. Consequently, dense cores are often identified and defined in terms of the specific tracer and molecular transition used in the observations.

Each molecular tracer is excited at different densities and temperatures, delineating structures with varying extents  \citep[e.g.,][]{Shirley2015_critdens, Andre}. As a result, there is no single tracer or threshold density that can be used to distinguish all different substructures within MCs. The criteria used to define certain regions (filaments, clumps, and cores) are thus conventional but ambiguous. In reality, MCs represent a density continuum, with substructures defined by the chosen observed tracer being separated for convenience, as these substructures do not correspond to distinct physical entities.

Some of the key observables of star formation in MCs that theoretical models aim to reproduce are well constrained by observations. One of them is the star formation efficiency (SFE), defined as
\begin{equation}
    \mathrm{SFE}(t) = \frac{M_{\star}}{M_{\mathrm{C}} + M_{\star}}, \label{eq:isfeeq}
\end{equation}
where $M_{\star}$ is the mass in stars within the region of study and $M_{\mathrm{C}}$ denotes the cloud mass. The SFE is generally thus interpreted as the fraction of a cloud's mass that is deposited in stars, and is generally a function of time. On the scales of entire GMCs, the SFE is observed to be $\sim 2\%$ \citep[e.g.,][]{Krumholz+19}, whereas in dense massive-star forming regions SFE$\sim$30-50\% \citep[e.g.,][]{Lada+03}. On the other hand, the  ``efficiency per free-fall time'', $\eff$, defined as
\begin{equation}
    \eff = \frac{ \langle \dot{M_{\star}} \rangle}{M_{\mathrm{C}}} \tau_{\rm ff},  \label{eq:sfrff_eq}
\end{equation}
where $\tauff$ is the mean free-fall time of the gas, and $\langle \dot{M_{\star}} \rangle$ is the average star formation rate (SFR), was introduced by \cite{sfrff} as a measure of the fraction of a cloud's mass that is transformed into stars over a global free-fall time, and thus as a measure of the actual efficiency of star formation. Note, however, that interpreting $\epsilon_{\rm ff}$ as an efficiency is problematic, as discussed by \cite{ZA+2025}. A more physically meaningful interpretation is that of a dimensionless ratio between the star formation rate (SFR) and the gas free-fall rate. Nevertheless, we retain the conventional terminology throughout this paper for consistency with the existing literature.

Generally, at scales of MCs there is consensus on a canonical value of $\eff \simeq 0.006-0.026$ with factor of 4 variation from cloud to cloud for resolved, nearby star-forming regions \citep[e.g.,][]{Pokhrel+21}. However, a dispersion of at least one order of magnitude exists both above and below this range \cite[see, e.g.,][]{Krumholz+19}. At the scale of dense cores, \citet{Louvet+14} studied the mini-starburst region W43-MM1 and found that $\epsilon_{\rm ff}$ increases toward the densest regions of the system, from $\sim 0.09$ up to $\sim 1.5$. Over the same range, the virial parameter decreases from $\alpha_{\rm vir}\simeq0.74$ to $\simeq0.11$, showing a continuous anticorrelation between $\epsilon_{\rm ff}$ and $\alpha_{\rm vir}$. While turbulence-regulated star formation models predict that $\epsilon_{\rm ff}$ should increase as $\alpha_{\rm vir}$ decreases, they generally anticipate this trend to weaken and approach a saturation regime at sufficiently low virial parameters \citep[e.g.,][]{sfrff, Padoan_Nordlund2011, Hennebelle_Chabrier2011}. The continued rise of $\epsilon_{\rm ff}$ reported by \citet{Louvet+14} down to $\alpha_{\rm vir}\simeq0.11$ therefore appears stronger than predicted by these models. Note also the fact that, as defined, $\eff$ can be larger than unity, which is one of the problems of interpreting it as an efficiency \citep{ZA+2025}.

Different theoretical models attempt to explain the values and trends of $\eff$, in particular the turbulent support and global hierarchical collapse (GHC) models \citep[see, e.g.,][]{Federrath15}. Numerical simulations of self-gravitating clouds achieve observationally consistent values of $\epsilon_{\rm ff}$ ($\sim 0.04$) either through the action of supersonic MHD turbulence \citep[e.g.,][]{Padoan+12} or by including stellar feedback processes, such as protostellar outflows \citep[e.g.,][]{Federrath15, Kim+21}. Other models, where turbulence decays while gravity drives the evolution collapsing the cloud and forming stars, produce reasonable values of $\eff$ when considering feedback effects (such as outflows, winds and ionizing radiation) \citep[e.g.,][]{Grudic2}. Furthermore, recent studies within the GHC framework report low values of $\epsilon_{\rm ff}$ \citep{BP+23, ZA+2025}. 

 In addition to feedback, gas accretion plays a crucial role in the evolution of $\epsilon_{\rm ff}$. Recent studies have highlighted that cores are not isolated systems but are continuously accreting material from their environment. It has been shown that this process can maintain low efficiency values even in the absence of strong feedback mechanisms, as the accretion onto star-forming clumps continually replenishes the gas supply, effectively keeping the conversion of gas into stars inefficient per unit of free-fall time \citep{alejandro_2020,Vazquez_Semadeni+19}.  Furthermore, accretion driven by hierarchical collapse has been argued to sustain turbulence and regulate the SFR \citep{ZA+2025},  implying that the detailed history of how a core gains mass is fundamental to understanding the observed values of $\epsilon_{\rm ff}$. 

The physical nature of $\eff$ is currently being extensively discussed. This work aims to investigate its origin and evolution in dense cores through numerical experiments of simplified spherical gravitational collapse, focusing on the fundamental physics (gravity) influencing this parameter behaviour in dense cores. Our objective is not to replicate the exact observational $\epsilon_{\rm ff}$ values in our simulations, but rather to examine how variations in accretion and core definition affect this parameter. Our paper is structured as follows. In Section \ref{analytical_stuff}, we present an analytical estimate of $\epsilon_{\rm ff}$ for a spherical core with a radial density profile. In Section \ref{sec:numerics}, we describe the numerical model used in this work and in Section \ref{sec:results} we present our results. Section \ref{sec:discussion} provides a discussion of these results. Finally, in Section \ref{sec:summary} we summarise our main findings and present the conclusions of this study.

\section{Analytical estimation of \texorpdfstring{$\epsilon_{\mathrm{\lowercase{ff}}}$}{epsilon\_ff}}
\label{analytical_stuff}

Is important for the purpose of this work to understand the spatial dependence of $\epsilon_{\rm ff}$. In general, it is well-accepted that the density profile for a collapsing system behaves as \citep{larson+69}

\begin{equation}
\rho(r) = \rho_0 \left(\frac{r}{r_0}\right)^{-p},
\end{equation}
where $\rho_0$ represents the gas density at a reference radius $r_0$, and $p$ is the density profile exponent, which is specific to the system and determined by the initial conditions or observational constraints. To explore the spatial dependence of the free-fall time $\tau_{\rm ff}$,
\begin{equation}
\tau_{\rm ff} = \sqrt{\frac{3 \uppi}{32 G \rho}},
\end{equation}
we first write it in terms of the density profile as
\begin{equation}
\tau_{\rm ff}(r) = \sqrt{\frac{3 \uppi}{32 G \rho_0} \left(\frac{r}{r_0}\right)^p}.
\end{equation}
Next, the total mass of the collapsing system, assuming it is spherically symmetric, is given by {$M_\mathrm{core} = \int_0^r 4 \uppi r'^2 \rho(r') \dif r'$}, which, using the density profile, becomes

\begin{equation}
M_{\mathrm{core}} = \frac{4 \uppi \rho_0 r_0^3}{3-p}
   \left(\frac{r}{r_0}\right)^{3-p} .
\end{equation}

\noindent

Finally, the star formation efficiency $\epsilon_{\rm ff}$ is related to the SFR, the core mass, and the free-fall time. Since the SFR is independent of the core radius, we can express $\epsilon_{\rm ff}$ as
\begin{align}
\epsilon_{\rm ff} &= \frac{\dot{M}_{\star}}{M_{\mathrm{core}}} \tau_{\rm ff} =
\frac{\dot{M}_{\star} (3-p)}{4 \uppi \rho_0 r_0^3} \left( \frac{r}{r_0} \right)^{p-3}
\sqrt{\frac{3 \uppi}{32 G \rho_0} \left( \frac{r}{r_0} \right)^p}  \nonumber \\
&= \frac{(3-p) \dot{M}_{\star}}{4 \uppi \rho_0 r_0^3} \sqrt{\frac{3 \uppi}{32 G \rho_0}}
\left( \frac{r}{r_0} \right)^{\frac{3}{2}p - 3} \label{eq:eff_final}
\end{align}

Note that $\epsilon_{\rm ff}$ is quite sensitive to the value of $p$, and for the case $p = 2$, $\epsilon_{\rm ff}$ becomes independent of radius \citep{ZA+2025}

\section{Numerical Methods} \label{sec:numerics}

For our numerical experiments, we use the Eurelian adaptive mesh refinement (\texttt{AMR}) FLASH v4.3 code \citep{FLASH1, FLASH2, FLASH3}. 
To solve the hydrodynamic equations, we employ the HLL3R solver \cite[][]{mhdmethod, MUSCL2}. Our models are isothermal and include physical processes, such as self-gravity \cite[Tree-based solver algorithm,][]{bhtree} and sink particles \cite{sink_particles_modulus}.

\subsection{Simulation Parameters} \label{parameters}

To understand the differential collapse of a dense core from an evolutionary perspective, we employ a three-dimensional idealized numerical model. This model consists of a numerical box of size $L_x = L_y = L_z = L_{\mathrm{box}} =4 $ pc, filled with cold molecular gas with a uniform background uniform density $\rho_0$. 
We impose a Gaussian overdensity at the center of the numerical box ($\bm{r}^{'}_0$) as the seed of a spherical collapse. Thus, the mass density in each cell at position $\bm{r}_{\{i,j,k\}}$ is given by
\begin{equation}
    \rho_{\{i,j,k\}} = \rho_0 + \rho_0 \mathrm{e}^{-\frac{(r_{\{i,j,k\}}-r_0)^2}{2\sigma^2}}, \label{eq:gauss_overden}
\end{equation}
\noindent
where $\sigma$ is the width of the Gaussian profile, chosen in direct relation to the size of the numerical box, $\sigma = L_{\mathrm{box}} / 5$. Our simulations are isothermal, with $T_0 = 10$ K, resulting in a sound speed of $c_s = 0.189 \, \kms$. Note that our initial conditions are similar to those used in \citet{Naranjo-Romero_2015}.

\begin{table}
	\centering
 \begin{tabular}{|c||c|c|c|c|}

	\toprule
	\multirow{2}{*}{\textbf{Parameters}}& \multicolumn{4}{c}{\textbf{Simulations}} \\  \cline{2-5}
	& C100& O100 & C1000 & O1000\\ \midrule \midrule

 	$n_0$ [cm$^{-3}$] $^*$&  \multicolumn{2}{c}{82.46} &\multicolumn{2}{|c|}{824.65} \\ \midrule
 	$\langle n \rangle$ [cm$^{-3}$]& \multicolumn{2}{c}{100}&  \multicolumn{2}{|c|}{1000}\\ \midrule
 	$\langle \tau_{\rm ff} \rangle$ [Myr] & \multicolumn{2}{c}{3.38859} & \multicolumn{2}{|c|}{1.07157}\\ \midrule
  	$M_T$ [M$_{\odot}$] & \multicolumn{2}{c}{300} & \multicolumn{2}{|c|}{3000}\\ \midrule
   	$M_J$ [M$_{\odot}$] & \multicolumn{2}{c}{60.81} & \multicolumn{2}{|c|}{19.23}\\ \midrule
   	$M_T/M_J$  & \multicolumn{2}{c}{4.933} & \multicolumn{2}{|c|}{156.006}\\ \midrule \midrule
 	$l_{\mathrm{max}}$ $^*$ & \multicolumn{2}{c}{6}&  \multicolumn{2}{|c|}{8}\\ \midrule
    $\Delta x$ [pc] & \multicolumn{2}{c}{0.016} & \multicolumn{2}{|c|}{0.0039}\\ \midrule
	$r_{\text{acc}}$ [pc] & \multicolumn{2}{c}{0.039} & \multicolumn{2}{|c|}{0.0098}\\ \midrule 
    $n_{\text{sink}}$ [cm$^{-3}$]& \multicolumn{2}{c}{4.715$\times10^5$}&  \multicolumn{2}{|c|}{7.544$\times10^6$}\\ \midrule\midrule
	Boundary $^*$ & \multirow{2}{*}{Closed} & \multirow{2}{*}{Open} &  \multirow{2}{*}{Closed} & \multirow{2}{*}{Open}\\
	Type& & & &\\
	
	\bottomrule
    \multicolumn{5}{l}{\footnotesize{$^*$ Input Parameters.}}\\
 
 \end{tabular}
	
	\caption{Parameters from the four simulations performed in this work. The columns correspond to specific simulations, and within each row, different important parameters are presented for the characterization of every simulation. From top to bottom, $n_0$ corresponds to the initial background uniform density, $\langle n \rangle$ is the initial mean density, $\langle\tau_{\rm ff} \rangle$ is the gravitational free-fall time, $M_T$ is the total initial mass within the numerical box, $M_J$ is the Jeans Mass at the beginning of the simulation, $M_T / M_J$ is the total initial mass and Jeans mass ratio, $l_{\mathrm{max}}$ is the highest refinement level possible in the simulation, $\Delta x$ is the minimum cell size in the box, $r_{\text{acc}}$ is the accretion radius of the sink particle, $n_{\text{sink}}$ is the critical density for a sink to be formed, and the \textit{Boundary Type} row tells the chosen BC.}
	
	\label{tab:parameters}
	
\end{table}

Four simulations were performed by varying the BCs and the initial uniform number density $n_0$.\footnote{We indistincly use  the mass density ($\rho$) or the number density ($n$), which are related by $\rho = \mu \, m_{\rm H} \, n$, where $\mu$ is the mean molecular weight and $m_{\rm H}$ the mass of the hydrogen atom. Throughout this work, we adopt $\mu = 2.3$, which is appropriate for molecular gas.} 
For this, we select two values for $n_0$, $82.46$ cm$^{-3}$ and $824.65$ cm$^{-3}$, so the mean number density at the start of the simulation $\langle n \rangle$, including the gaussian overdensity at the center of the numerical box, will be approximately $100$ cm$^{-3}$ and $1000$ cm$^{-3}$, respectively.

Table \ref{tab:parameters} shows the main parameters from each numerical experiments. 
The values presented in the table can be divided into three blocks:
\begin{enumerate}

    \item 
The first six rows are values related with the mean initial number density ($\langle n \rangle$), being $\langle\tau_{\rm ff} \rangle$ the gravitational free-fall time, which was calculated as
    \begin{equation}
        \langle\tau_{\rm ff} \rangle = \sqrt{\frac{3 \uppi}{32 G \langle n \rangle \mu m_{\rm H} }}. \label{eq:tff}
    \end{equation} 
    \item The next four rows correspond to numerical resolution parameters and the formation of sink particles. It is important to recall that since the threshold density for sink formation ($n_{\mathrm{sink}}$) depends on the maximum refinement level possible $l_{\mathrm{max}}$, and this has to be an integer number, there is no enough freedom to change $n_{\mathrm{sink}}$ to a specific desired value. The selection of $n_{\mathrm{sink}}$ was made so that the ratio between $n_{\mathrm{sink}}$ and $n_0$ would be the most similar possible between the four simulations.  Thus, although $\sim5\times10^4$~cm$^{-3}$ in the O100 and C100 simulations might seem a very low density threshold for sink formation (more comparable to typical clump densities, e.g., \citealt{RomanZuniga2026}), the numerical setup restricts us to do so.
    The parameters $\Delta x$ and $r_{\mathrm{acc}}$ correspond to the minimum cell size and the accretion radius of the sink, respectively. The $r_{\mathrm{acc}}$ is obtained as $r_{\rm acc} = 2.5 \times \Delta x$.

We use adaptative refinement so that the Jeans length in each cell is resolved by at least four grid cells \citep{truelove_criterion}.

    \item The last row shows the BCs selected for each simulation, with \textit{closed} referring to diode type BC, and \textit{open} to outflow type BC, as we explain below.
\end{enumerate}

\subsection{Boundary Conditions}

In this study, we utilize two types of BCs: \textit{open} and \textit{closed}.
The first BC enforces a zero-gradient condition at the boundaries, maintaining the fluid variables (e.g., density, velocity, and pressure) at the same values as the nearest corresponding cell within the domain. In contracting systems, this type of BC facilitates material inflow to prevent empty cells at the boundaries.

The closed BCs is analogous to the \textit{outflow} type, but it restricts gas flow into the domain. If necessary, the normal velocity components in guard cells are set to zero.

For the purpose of this work, we employ open and closed BCs (hereafter OBC and CBC, respectively). OBC simulates a scenario where the collapsing system is continually supplied with mass from an external reservoir, indicating that the system is not isolated. On the other hand, CBC is equivalent to an isolated collapse, and the system has a fixed material reservoir to accrete onto the core and sink \citep{Flash_manual}.

\subsection{Definition of the Core}

We consider three different density thresholds to define the boundary of the core and explore the impact of these thresholds on the measured star formation efficiency. The selected density thresholds are $n_{\mathrm{thr}} = 10^3$ cm$^{-3}$, as would roughly correspond to low-density clumps, $n_{\mathrm{thr}} = 10^4$ cm$^{-3}$, corresponding to the transition between clumps and cores, and $n_{\mathrm{thr}} = 10^5$ cm$^{-3}$, representing high-density cores.

The criteria used to define our structures are motivated by observational data. Different molecular gas species emit from distinct regions based on the density and temperature necessary for collisional excitation and radiation emission \citep[e.g.,][]{core/clump/fil, Shirley2015_critdens}. For instance, the most common low-density tracer for clumps is the $^{13}$CO molecule.
Conversely, low-level transitions of NH$_3$ and N$_2$H$^+$ serve as effective high-density tracers for identifying dense cores, as they remain detectable at densities where other molecules are typically frozen onto dust grains \citep[e.g.,][]{Pillai_2006, Busquet2011_N2Hp-depletion}. However, the adopted density thresholds should be interpreted as idealized proxies for observational tracers rather than direct equivalents.

\section{Results} \label{sec:results}

Four simulations were performed. All simulations are gravitationally unstable and exhibit the expected behaviour of spherically symmetric collapse, with mass flowing toward the centre of the numerical box and collapsing from the outside in \citep[e.g.,][]{whitworth_85,Gomez_2021}. Eventually, a sink particle forms at the core centre. For clarity, we will refer to the O100 and C100 models as the low-mass or low-density models/cores, while the O1000 and C1000 models will be termed the high-mass or high-density models/cores, with \textit{O} denoting models with OBCs and \textit{C} indicating models with CBCs.

\subsection{General Evolution} \label{sec:general_evolution}

\begin{figure*}
	\centering
	\subfigure[C100 simulation.]{\includegraphics[width = 0.45\textwidth]{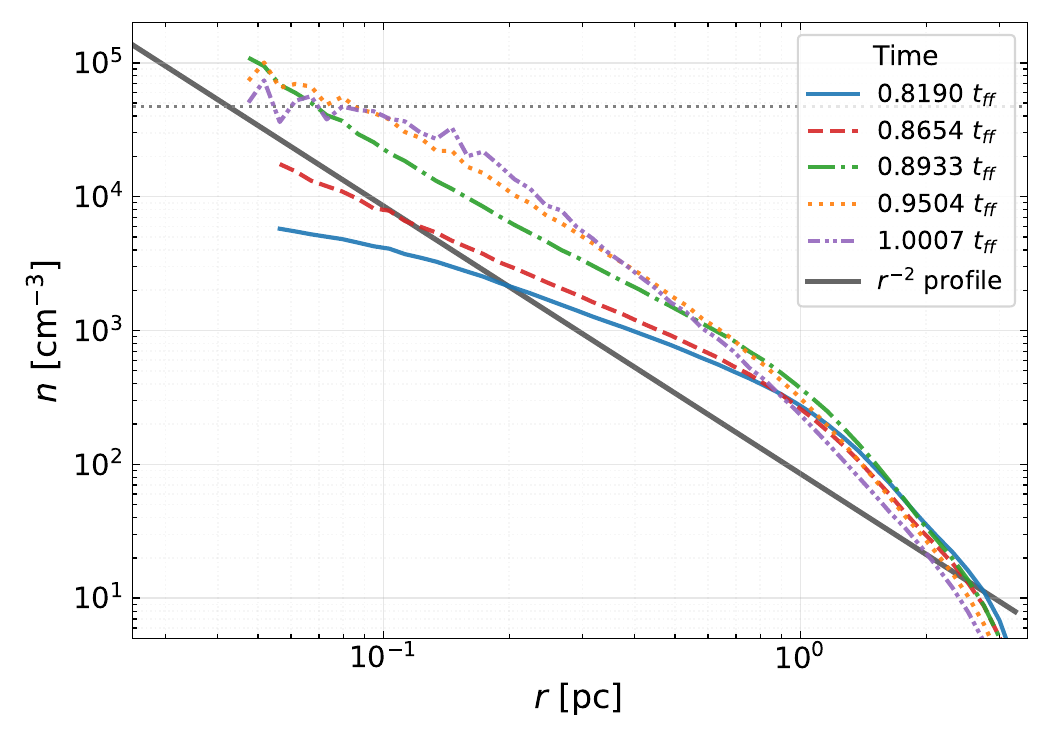}}
		\hfil
   	\subfigure[O100 simulation]{\includegraphics[width = 0.45\textwidth]{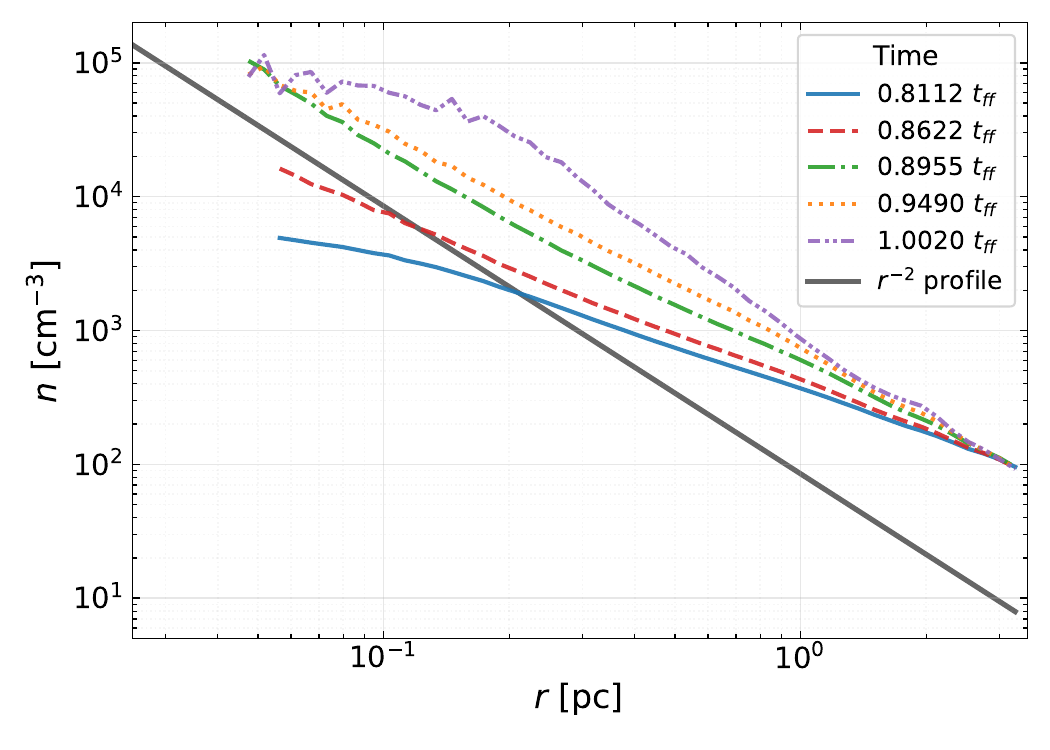}}

    \subfigure[C1000 simulation]{\includegraphics[width = 0.45\textwidth]{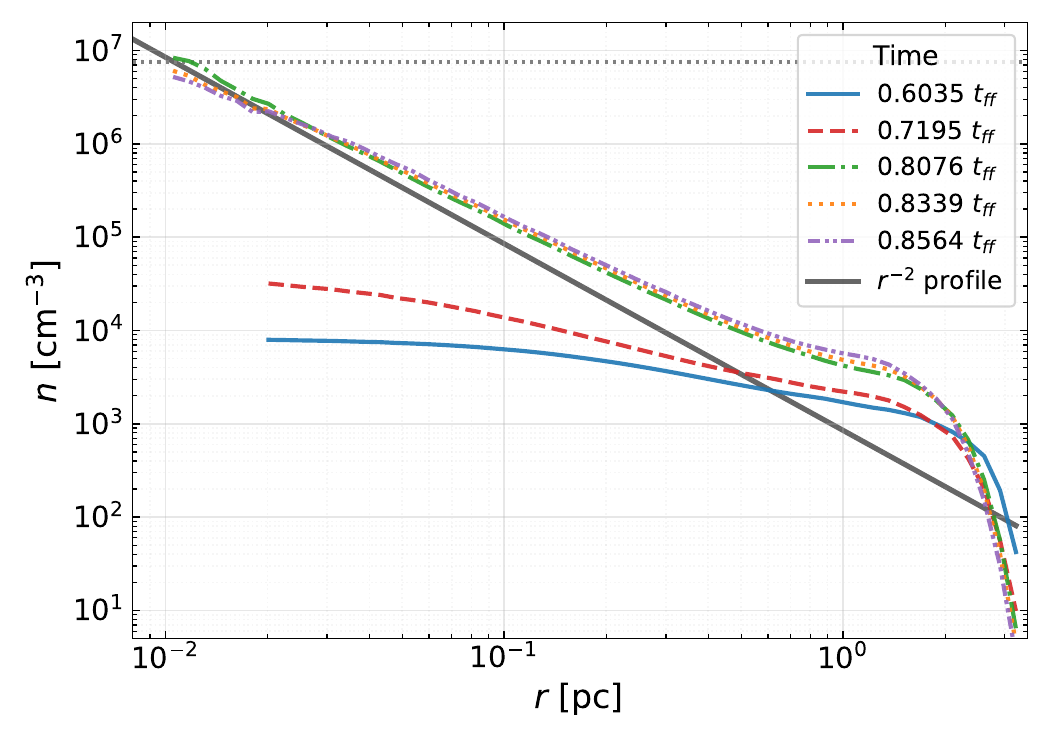}}
	\subfigure[O1000 simulation]{\includegraphics[width = 0.45\textwidth]{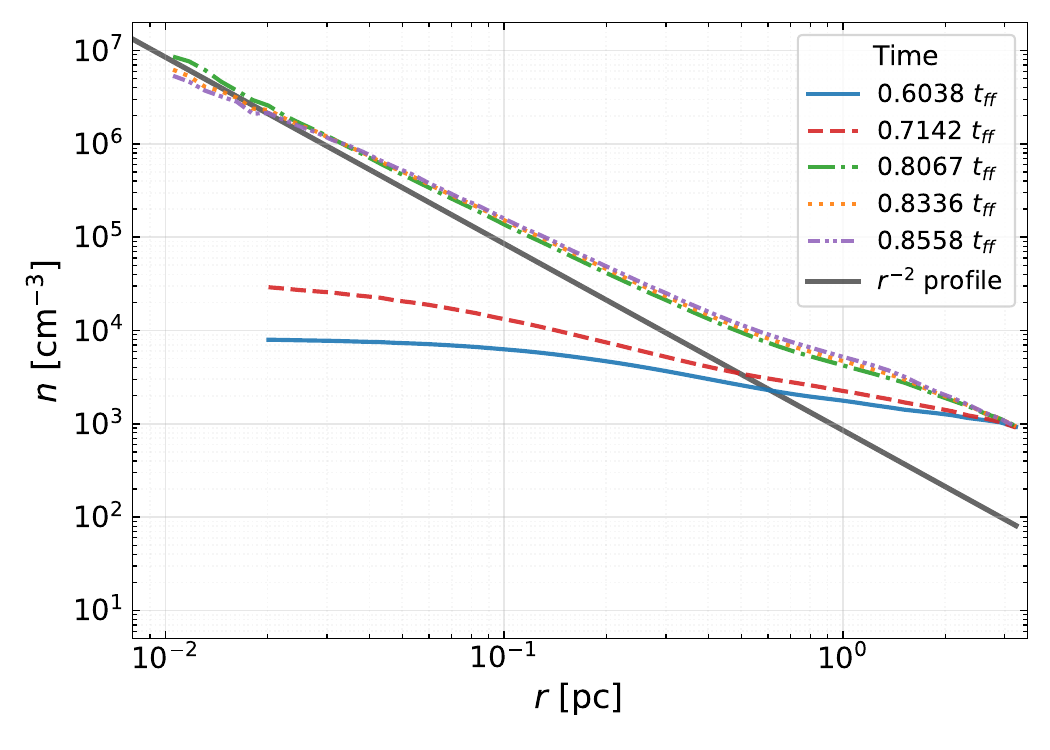}}

	\caption{The density profile (measured as the mean density in thin annuli at each radius) for each simulation is presented, highlighting five different moments in time: one immediately after sink formation (green dot-dashed lines), along with two time points preceding and two following this time. Each panel shows an $r^{-2}$ profile (solid black line), with a dashed horizontal line indicating the threshold number density for sink formation ($n_{\mathrm{sink}}$). The \textit{top left panel} corresponds to the C100 simulation, \textit{top right panel} to O100, \textit{bottom left panel} to C1000, and \textit{bottom right panel} to O1000.}
	\label{fig:dens_prof}
\end{figure*}

The chosen initial Gaussian overdensity amplitude is moderate---only a factor of two relative to the background density (Eq.~\ref{eq:gauss_overden}). Thus, as collapse proceeds, the cores develop a power-law density profile. Figure \ref{fig:dens_prof} shows the temporal evolution of the number density profile for each simulation.\footnote{ Radial density profiles were computed by spherically averaging the gas density within concentric shells centred on the box centre.} Each panel presents profiles for five distinct times around the sink-formation time ($t_{\star} = 0.8933 \; t_{\mathrm{ff}}, \; 0.8955 \; t_{\mathrm{ff}}, \;0.8076 \; t_{\mathrm{ff}}, \;$ and $0.8067 \; t_{\mathrm{ff}}$, or the C100, O100, C1000, and O1000 models, respectively, indicated by the green line). Note that $t_{\star}$ is the time of sink particle creation, measured in units of the simulation's $t_{\mathrm{ff}}$. An $ r^{-2}$ reference profile is overplotted, together with a dashed horizontal line indicating the sink-formation threshold density, $n_{\mathrm{thr}}$, for each run. The figure highlights several features. Density profiles were truncated in models with closed boundary conditions (CBCs), whereas runs with open boundary conditions (OBCs) showed material replenishment from the box boundaries. In all models, the profile initially exhibited a relatively flat slope; it then steepened and approached an approximately constant value near $t_{\star}$.

From Figure \ref{fig:dens_prof}, we measured the best-fitting power-law density slope at the sink-formation time, $t=t_\star$, over the full radial range shown. We obtained values of $2.05 \pm 0.04$, $1.99 \pm 0.03$, $1.76 \pm 0.01$, and $1.75 \pm 0.01$ for the C100, O100, C1000, and O1000 models, respectively.
These values are consistent with with the numerical results of \citet{Naranjo-Romero_2015}, who showed that prestellar collapse proceeds from large to small scales (i.e. in an outside-in manner), while the density profile steepens progressively toward a Larson--Penston-like configuration with an outer slope approaching $p\simeq2$. They are also consistent with the analytical model of \citet{Gomez_2021}, in which collapsing cores evolve through a sequence of increasingly steeper density profiles and converge toward the attractor solution $p=2$. 

The sink particle formed earlier in the higher density models (C1000 and O1000)  than in the lower density experiments (C100 and O100). This difference reflects the greater dynamical importance of thermal pressure in the lower-density models, where the ratio of thermal to gravitational energy is larger, delaying the collapse and the onset of sink formation. For all the models, the initial sink mass was small ($ 2 $--$ 3~M_\odot $), with the majority of its mass acquired subsequently through accretion.

\begin{figure}
    \centering
    \includegraphics[width = 0.45\textwidth]{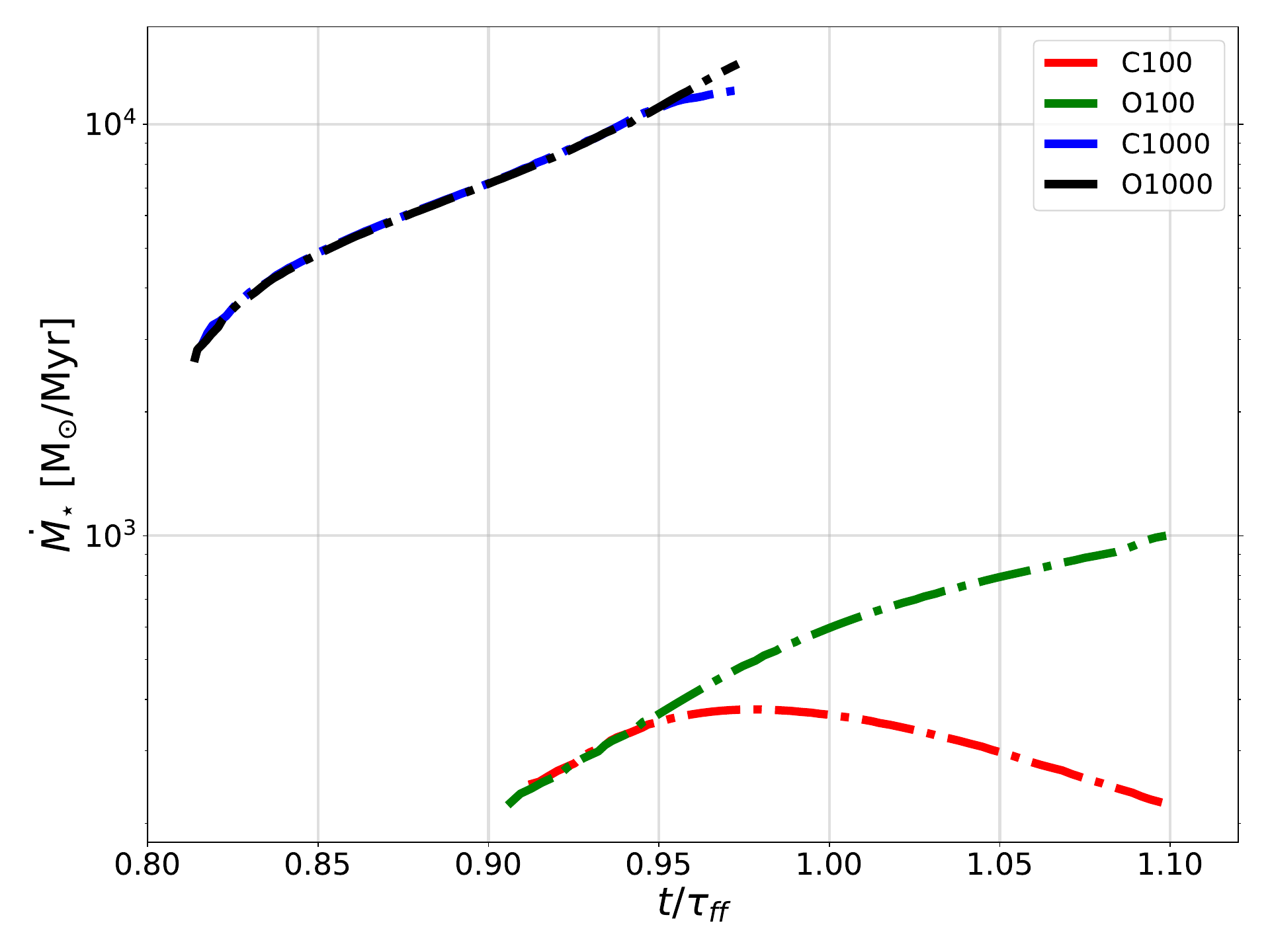}
    \caption{Evolution of the star formation rate ($\dot{M}_{\star}$) for each simulation, which is represented with different colors: red for the C100 simulation, green for the O100 simulation, blue for the C1000 simulation and black for the O1000 simulation.}
    \label{fig:sink_acc}
\end{figure}

\begin{figure*}
    \centering
    \includegraphics[clip,width =0.95\textwidth]{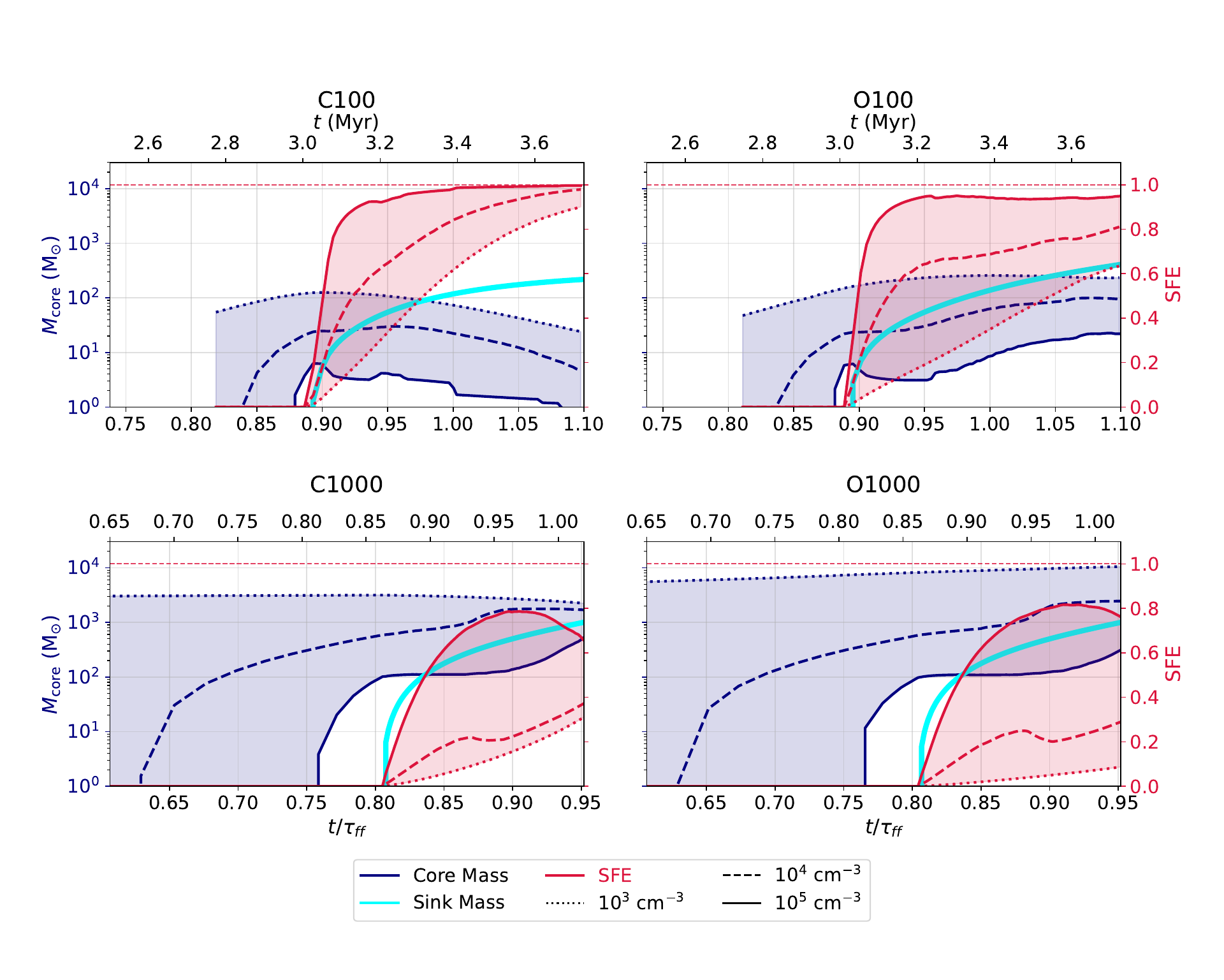}
    \caption{ Core mass and SFE evolution (left and right vertical labels, respectively) for each simulation. Here, we show three different core definitions as follows: the core with $n_{\mathrm{thr}} = 10^3$ cm$^{-3}$ with dotted lines, the core with $n_{\mathrm{thr}} = 10^4$ cm$^{-3}$ with dashed lines, and the core with $n_{\mathrm{thr}} = 10^5$ cm$^{-3}$ with solid lines . The \textit{top left panel} corresponds to the simulation labeled as C100, the \textit{top right panel} to O100, the \textit{bottom left panel} to the C1000, and the \textit{bottom right panel} to the O1000 model. The shaded blue region represents the range of core masses within the three threshold densities used for defining the cores, and the red region the corresponding SFE range, respectively. To better understand the evolution, both the time in Myr (top x-axis) and in $\tau_{\rm ff}$ units (bottom x-axis) are shown.
    }
    \label{fig:core_dependent}
\end{figure*}

\subsection{Star Formation Rate and Efficiency} \label{subsec:SFE}

We computed the instantaneous SFR ($\equiv \dot{M}_{\star}$; see Fig.~\ref{fig:sink_acc}), i.e., the sink accretion rate as a function of normalized time ($t/\tau_{\rm ff}$). Interestingly, at early times the SFR accelerates for all models \footnote{ The SFR traces a roughly straight line in Fig.~\ref{fig:sink_acc}, which uses a log–linear axis, suggesting exponential growth.}, although the C100 model shows this behavior only during the very initial stages. The more massive cores (models C1000 and O1000; blue and black lines) reach SFRs more than an order of magnitude higher than those in the low-density models (C100 and O100; red and green lines).

 On the other hand, parameters such as the SFE and $\epsilon_{\rm ff} $ depend not only on the evolutionary state of the core, but also on the specific definition of the core. Fig.~\ref{fig:core_dependent} shows the evolution of the core mass (blue curves), the sink mass (cyan curves), and the instantaneous SFE (red curves; Eq.~\ref{eq:isfeeq}) for the cores defined by the threshold densities $ n_{\mathrm{thr}} = 10^3 $, $ 10^4 $, and $ 10^5 $~cm$^{-3}$ (solid, dashed, and dotted lines, respectively).

We can identify several trends from the figure. First, the O1000 and C1000 models formed significantly more massive cores than the O100 and C100 models, irrespective of the density threshold adopted. Second, while the sink mass grew in an accelerated way in all models, the evolution of the core mass depended on both the boundary conditions and the adopted core definition. In low-density models (C100 and O100), the core defined using the lowest density threshold ($n_{\rm thr}=10^3$ cm$^{-3}$) either decreased after sink formation (C100) or increased moderately due to gas replenishment (O100). As a consequence, the SFE rose extremely rapidly in model C100, reaching values close to unity, whereas in model O100 the SFE grew more slowly and remained below unity. In high-density models (C1000 and O1000), the core mass remained large and relatively stable even for the lowest density threshold, yielding SFE values of $\mathrm{SFE} \lesssim 0.4$. For the highest threshold ($n_{\rm thr}=10^5$ cm$^{-3}$), the core became more compact and its mass became comparable to that of the sink, resulting in higher SFE values ($\mathrm{SFE} \lesssim 0.8$).

However, note that the SFE approached unity in low-density models (C100 and O100). This behaviour reflects the fact that these simulations reached a highly evolved stage of collapse, in which a large fraction of the available gas had already been accreted onto the sink particle, and should therefore not be interpreted as evidence that lower-density environments are intrinsically more efficient at forming stars.


\subsection{\texorpdfstring{$\epsilon_{\rm ff}$}{epsilon\_ff} as the ratio of the SFR to the gas-infall rate} \label{subsec:SFEff}

We have also measured $\epsilon_{\rm ff}$ (Eq.~\ref{eq:sfrff_eq}) for all our models. For this, we used the time-averaged, $\langle {\rm SFR} \rangle$, computed as the sink mass divided by the elapsed time since sink formation.\footnote{ {\rm We adopted a time-averaged rather than an instantaneous SFR because observational estimates of star formation rates are generally inferred by averaging over finite time intervals associated with young stellar populations.} } The gas mass was measured for cores defined using different threshold number densities ($n_{\rm thr}$), and the corresponding free-fall time was calculated from the mean density of the gas enclosed within each threshold-defined core.

\begin{figure}
	\centering
	\includegraphics[width = 0.49\textwidth]{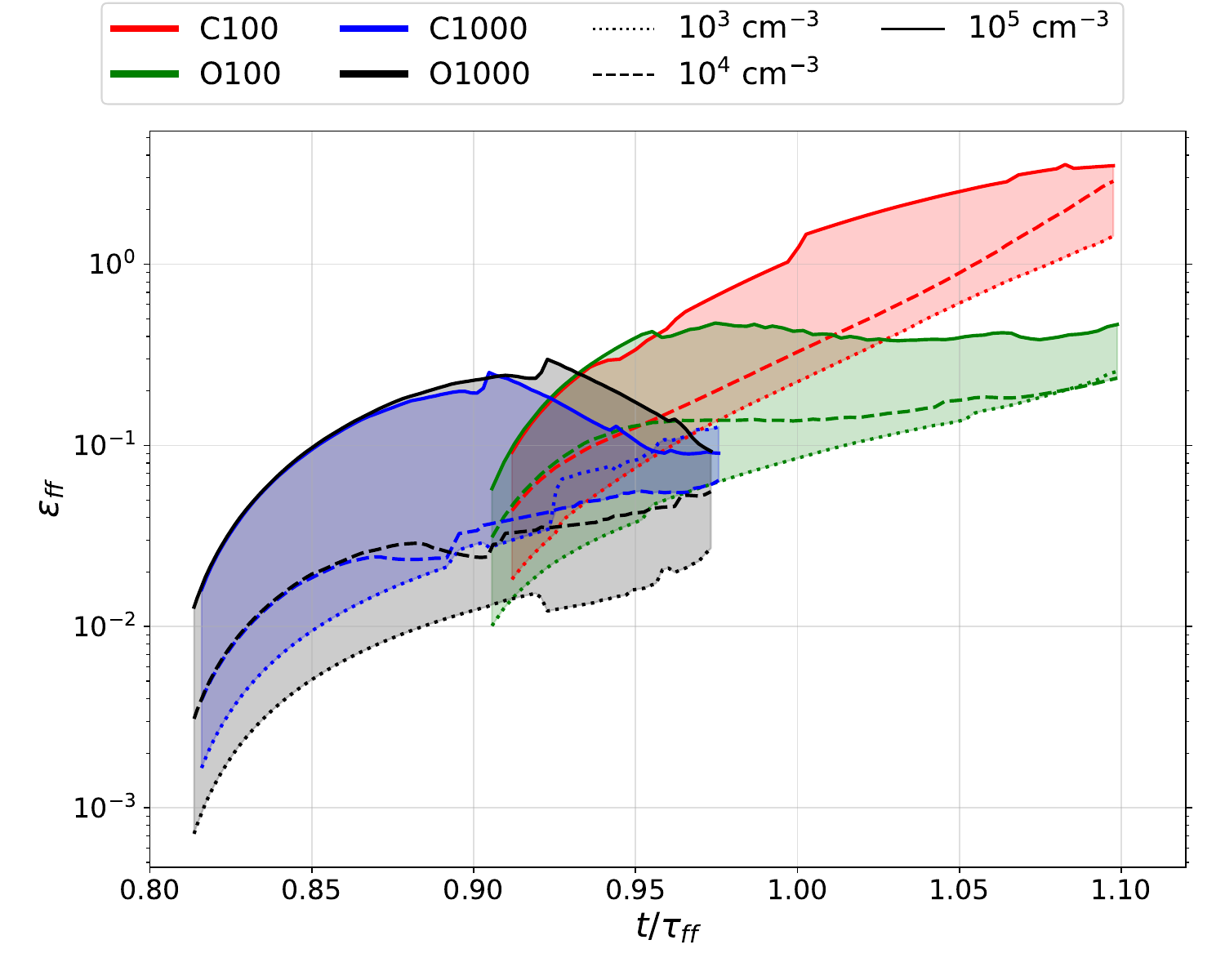}

	\caption{ Evolution of the ratio of the SFR to the gas-infall rate ($\epsilon_{\rm ff}$) for each simulation. As in Figure \ref{fig:core_dependent}, we show three core definitions as follows: the core with $n_{\mathrm{thr}} = 10^3$ cm$^{-3}$ with dotted lines, the core with $n_{\mathrm{thr}} = 10^4$ cm$^{-3}$ with dashed lines, and the core with $n_{\mathrm{thr}} = 10^5$ cm$^{-3}$ with solid lines. Each simulation is represented with each colored region: red for the C100 simulation, green for the O100 simulation, blue for the C1000 simulation and black for the O1000 simulation.}
	\label{fig:SFRff_evol}
\end{figure}

\begin{figure*}
	\centering
	\subfigure[C100 simulation]{\includegraphics[width = 0.49\textwidth]{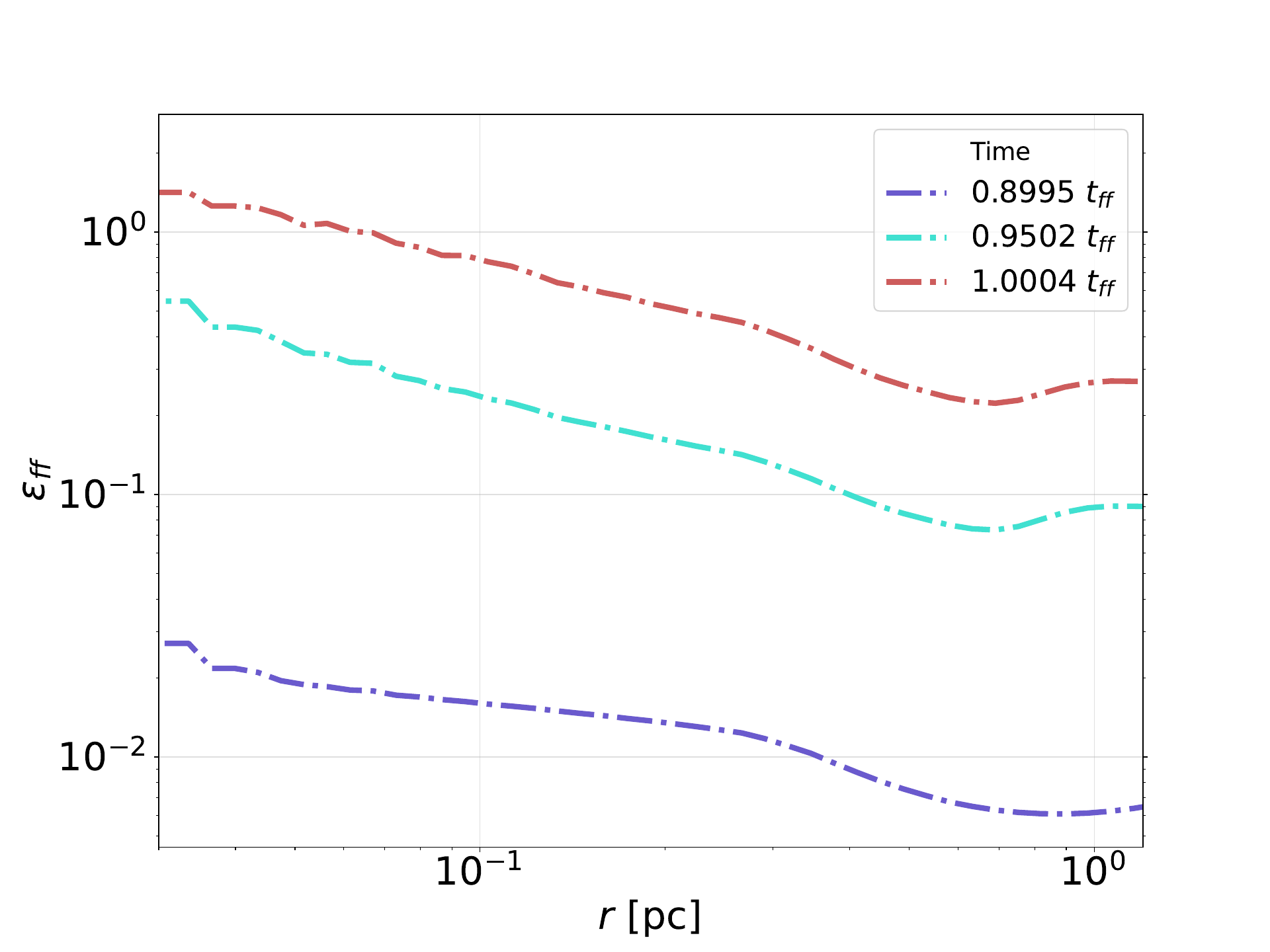}} \label{sfr_evol_103}
		\hfil
   	\subfigure[O100 simulation]{\includegraphics[width = 0.49\textwidth]{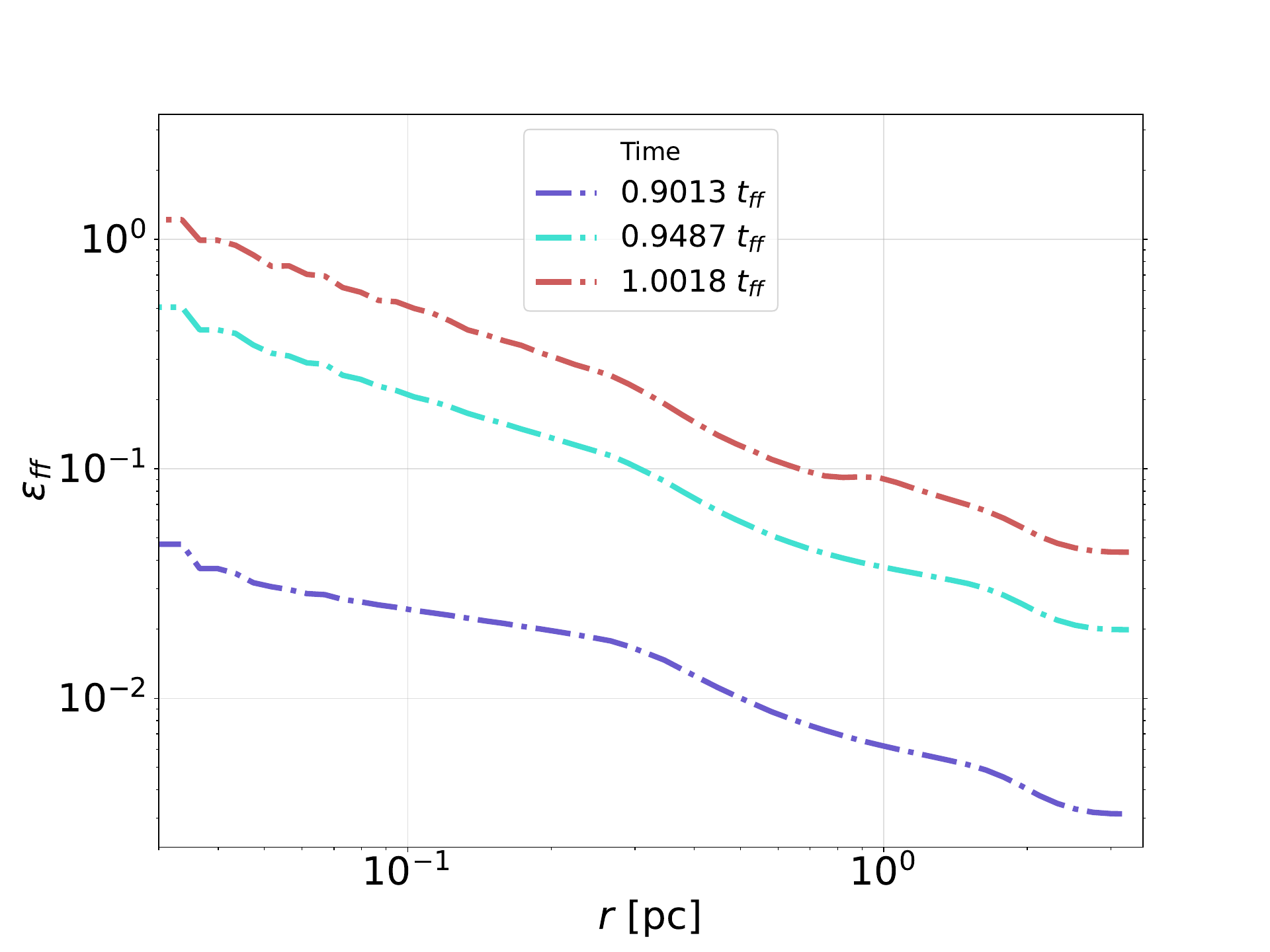}}

    \subfigure[C1000 simulation]{\includegraphics[width = 0.49\textwidth]{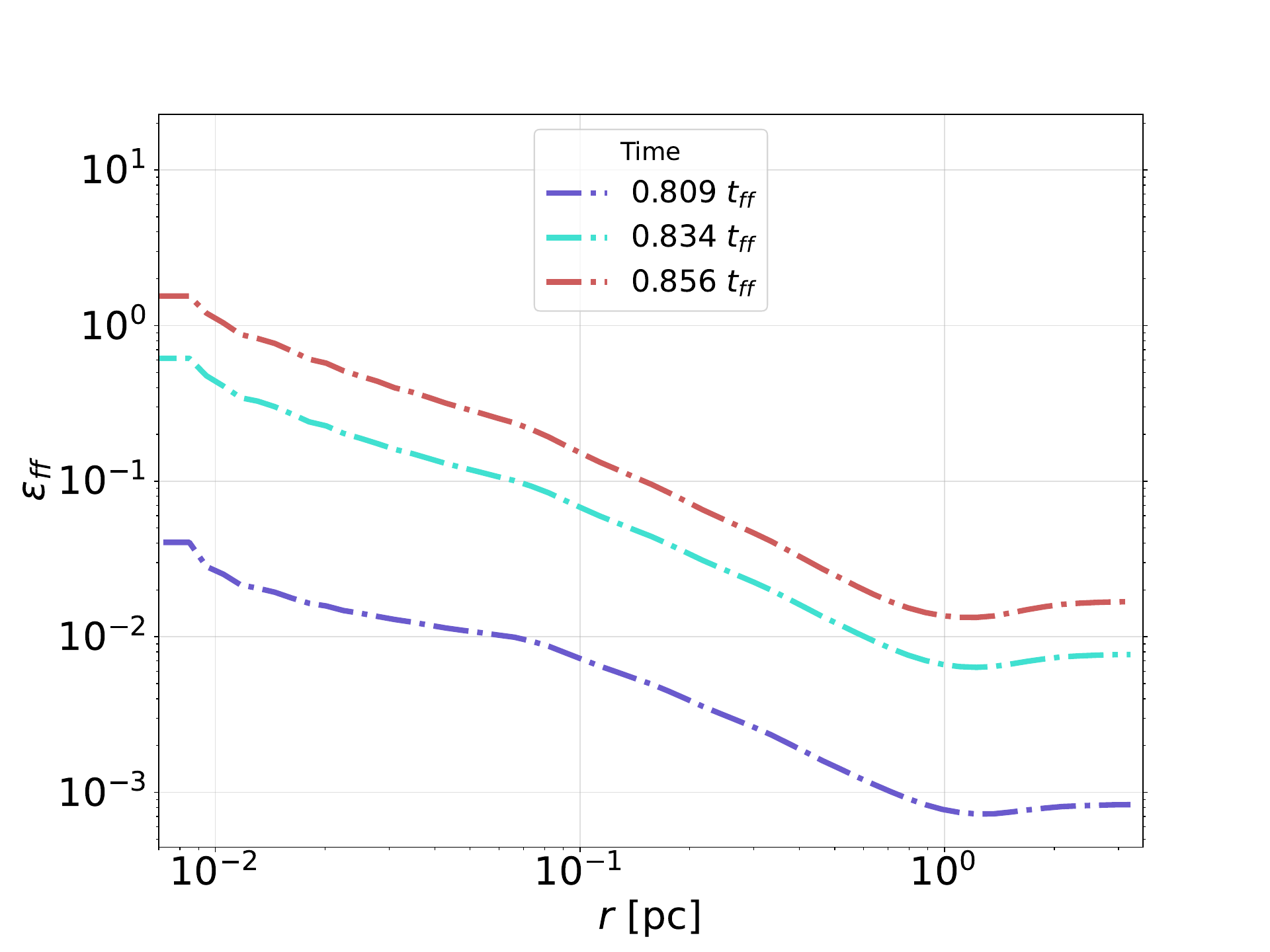}}
	\subfigure[O1000 simulation]{\includegraphics[width = 0.49\textwidth]{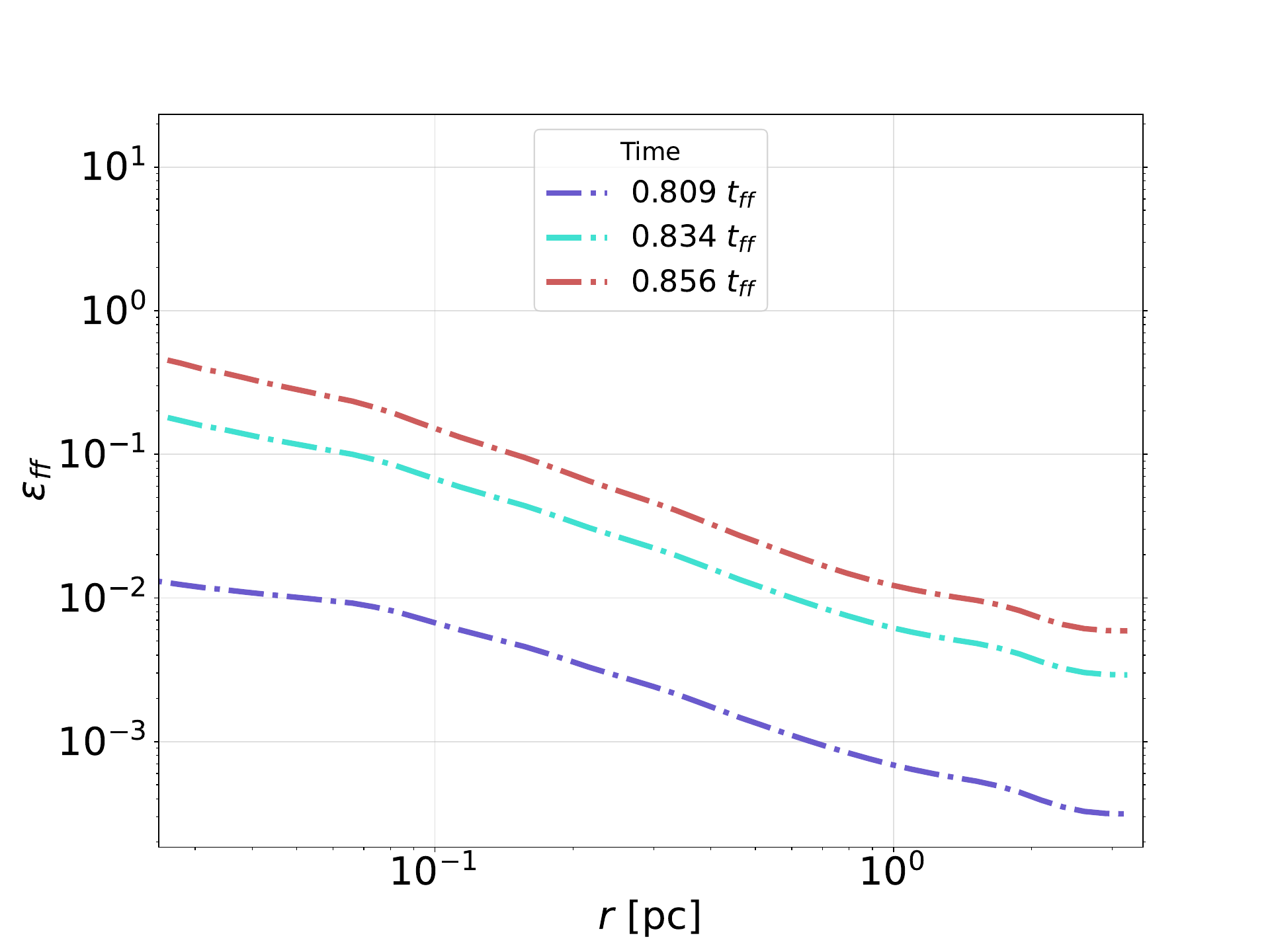}}

	\caption{ Profiles of the ratio of the SFR to the gas-infall rate ($\epsilon_{\rm ff}$) for each simulation. The \textit{top left panel} corresponds to the simulation labeled as C100, the \textit{top right panel} to O100, the \textit{bottom left panel} to the C1000, and the \textit{bottom right panel} to the O1000. The {\it dotted} colored lines depict the expected slopes for the indicated values of the density-profile slope, $p$.
    }
	\label{fig:SFRff_PROF}
\end{figure*}

The temporal evolution of $\epsilon_{\rm ff}$, shown in Fig.~\ref{fig:SFRff_evol}, shows a rapid increase immediately after sink formation, followed by more moderate growth or even decrease at later times. This behaviour is particularly evident in the high-density models (C1000 and O1000; blue and black lines in Fig.~\ref{fig:SFRff_evol}), where $\epsilon_{\rm ff}$ remains approximately constant over most of the evolution after the initial rise. Such behaviour suggests that, during this stage, the core gas infall rate ($M_{\rm core}/\tau_{\rm ff}$) roughly keeps pace with the growth of the $\langle {\rm SFR} \rangle$.  In contrast, for model C100, $\epsilon_{\rm ff}$ continues to increase with time, which is expected because the core mass (defined at all density thresholds) decreases and, without external gas replenishment, the gas infall rate drops. The core definition, based on different density thresholds ($n_{\rm thr}$), has a greater impact across all models, with $\epsilon_{\rm ff}$ tending to increase in models defined with higher density thresholds. This effect is essentially due to the fact that the stellar-to-gaseous mass ratio increases as the threshold is raised, because the sink particles are localized in a much smaller volume than that of their parent cores. This effect is also reflected in the radial profiles of $\epsilon_{\rm ff}$ (Fig.~\ref{fig:SFRff_PROF}).

Interestingly, $\epsilon_{\rm ff}$ is systematically higher in the low-mass cores than in the high-mass ones. This behaviour highlights the fact that $\epsilon_{\rm ff}$ depends not only on the SFR but also on the rate at which gas is supplied to the core. In the high-density models, the larger gas infall rates driven by the surrounding reservoir tend to compensate for the enhanced star formation activity, leading to lower values of $\epsilon_{\rm ff}$ despite their larger SFRs.

In Fig.~\ref{fig:SFRff_PROF}, we arbitrarily select three different times after the sink formation and plot the radial profile of $\epsilon_{\rm ff}$ for all models. As time progresses, the profile shifts towards higher $\epsilon_{\rm ff}$ values for all models, while preserving the shape of the profile. Radially, $\epsilon_{\rm ff}$ generally decreases with radius in all models, roughly following the expected trend based on the relationship $\epsilon_{\rm ff} \propto r^{3(\frac{p}{2} - 1)}$ (dotted lines in Fig.~\ref{fig:SFRff_PROF}; see Eq.~\ref{eq:eff_final}). However, in the high-density models, the decline is faster than predicted, probably reflecting a deviation from the power-law relationship in the density profile in the range $r\gtrsim 0.1 \, {\rm pc}$ (see Fig~\ref{fig:dens_prof}).

\section{Discussion} \label{sec:discussion}

\subsection{The dependence of \texorpdfstring{$\eff$}{epsilon\_ff} on the density profiles and the internal and external accretion rates}

 Our results from the previous sections indicate that, even in the absence of any forms of support, $\eff$ depends on the local conditions of the star forming region. We now discuss these dependencies.

To this end, we measure the density profiles of the cores, as well as properties related to sink formation. Figure \ref{fig:sink_acc} shows that the sink accretion rate increases with time, at least during the initial stages of evolution for all models, which is consistent with the behaviour observed in collapsing clouds \citep[e.g.,][]{ZA+2025} and with observations \cite[e.g.,][]{Stahler_Palla05}. The SFR in the C1000 and O1000 models is similar, while in C100, the accretion is less efficient than in O100. This difference arises from the open BCs in O100, which allow material injection into the box, providing a growing reservoir for the sink. The absence of this effect in C1000 is probably due to the fact that the mass of the numerical box in this case is much larger mass than the initial Jeans mass, and therefore there is plenty of material that can accrete onto the core despite the closed boundaries. 

On the other hand, as the density threshold ($n_{\mathrm{thr}}$, used to define a core) is increased, the SFE increases due to the cores becoming smaller and containing less mass, while the stellar mass near the density peak remains the same, independently of the threshold (compare, for example, the cyan solid line and the three dark blue lines in Fig.~\ref{fig:core_dependent}).

 It is noteworthy that, in Fig.\, \ref{fig:core_dependent}, we observe that the core mass in run C100 {\it decreases} over time, regardless of the threshold density, while the sink mass continues to increase. Instead, for the three other simulations, the core mass seems to approach a nearly constant value at the lowest threshold, while, at the largest threshold, it continues to increase on average. This result can be understood because run C100, with its closed boundary conditions, exhausts the gas in the core, thus reaching a very large SFE. Instead, the runs with open boundary conditions allow continuous replenishment of the core material, and thus the core mass can remain approximately constant, and the SFE remains smaller, especially at low $\nth$, as the core accretes nearly as much mass as it transfers to the sinks. The fact that a similar behavior is observed for the closed-boundary run C1000 can be understood because the total mass in this simulation is substantially larger than that in run C100, and therefore the fraction of it that goes into the core remains small, even with the closed boundary conditions, since the sink maintains a comparable accretion rate in both the OBC and CBC cases.

However, this effect becomes less noticeable as we increase the number density threshold ($n_{\mathrm{thr}}$) used to define the core, as seen in Figure \ref{fig:core_dependent}.
The fact that gas accumulation maintains a low SFE has also been acknowledged numerically by \citet{alejandro_2020} and, observationally, by, e.g., \cite{Zhang2025_ATOMS-lowSFE}.

Regarding $\epsilon_{\rm ff}$,  we note in Figure \ref{fig:SFRff_evol} that we measure a wide range of values for it, often much larger than the ``canonical'' value $\epsilon_{\rm ff} \approx 0.01$  \citep[e.g.,][]{Krumholz+19, Pokhrel+21}, and we obtain values up to $\sim 4$ for the C100 model. However, we should note that:

\begin{itemize}

\item The ``canonical'' value refers to regions on the scale of molecular clouds \citep[e.g.,][]{Lada+10,Pokhrel+21} or larger \citep[e.g.,] [] {Krumholz+19}, but values reported by \cite{Louvet+14} for a mini-starburst are in the range ($\epsilon_{\rm ff} \sim 0.09 - 1.5$).

\item Our idealized simulations have only one centre of collapse, whereas,  in reality, even cores contain several fragments that compete to accrete mass \citep[see, e,g.,][]{Palau+14, Xu2024_ASSEMBLE}.

\end{itemize}

Our idealized simulations are designed exclusively to investigate the effect of the core definition threshold and the role of accretion on $\eff$. Indeed, in \citet{ZA+2025} we showed that the presence of continuous accretion, together with a sufficiently steep density profile, can give rise to low measured values of $\eff$, because the size of the region where stars are forming ``instantaneously'' (i.e., on a much shorter timescale than the evolutionary timescale of the core or clump itself) is much smaller than the size of the whole region occupied by the gravity-driven collapse flow. However, since in general the cores are defined on practical rather than dynamical considerations \citep[although, see, e.g.,]{Gong2011, Mao2020}, the definition of the core itself affects the measured value of $\eff$. Also, since the presence of continued accretion onto the core is crucial in this interpretation, we also have investigated the effects of open versus closed boundaries.

In Figure \ref{fig:SFRff_evol} we compare the evolution of $\epsilon_{\rm ff}$ for our simulated cores defined with different density thresholds, and observe that $\epsilon_{\rm ff}$ takes lower values for the lower density thresholds. This is because the cores with lower density thresholds have more mass to accrete to the sink particle (as they are bigger in size, they have more material available) but the sink accretion rate is the same regardless of the definition of the core. This illustrates how the core definition can directly affect the $\epsilon_{\rm ff}$ (and SFE) measured within a region. This is important because in a star-forming region, the substructures are not isolated entities but part of a continuum, whose borders are arbitrarily defined either by observational or numerical biases.

 Analogously, the simulations with open boundaries (thus allowing continuous, albeit not fully physically correct, accretion), exhibit systematically lower $\eff$ than their closed-boundary counterparts, clearly demonstrating the fundamental role of accretion onto the core in maintaining a low efficiency, as well as its interpretation as the ratio of the accretion rate onto the stellar component to the theoretical collapse rate of the instantaneous mass reservoir.

Finally, in Fig.\ \ref{fig:SFRff_PROF} we plot the radial dependence of $\epsilon_{\rm ff}$ , for various times in each simulation, as resulting from the measured radial density profile of each simulated core, which in turn gives the radial profiles of $M_{\rm core}(r)$ and $\tauff(r)$ that enter eq.\ \eqref{eq:eff_final}. We notice a similar behaviour between the curves of the $\epsilon_{\rm ff}$ evolution in the four cores, so that all curves approach an $r^{-2}$ profile, as predicted in the analysis by \cite{Gomez_2021}. 
 In particular, the density profiles of the C100 and O100 simulations reach slopes very close to $-2$ at $t \approx \tauff$. The high-density simulations C1000 and O1000 do not, but this is most like due to the fact that these simulations had to be stopped at $t < \tauff$ for numerical reasons (cf.\ Sec.\ \ref{sec:limitations}). In any case, none of the simulations approach the radius-independent $\eff$ profile that would be expected from eq.\ \eqref{eq:eff_final} for a density profile slope $p=2$. This can be attributed to the fact that the radial density profile only approaches a power law in a limited range of radii. Nevertheless, $\eff$ is always seen to decrease with increasing radius, as a consequence of the localized nature of the star-forming sites, so that the SFR does not vary as the distance from them increases,\footnote{At least while a new star-forming site is encountered.} while the gas mass does. 

\subsection{Resolution Effects}

 It is important to determine if our cores are well resolved. For this, we determine whether the cores defined in our simulation are large enough to be resolved by the numerical grid. We do this by comparing the size of the cores to the accretion radius of the sink ($r_{\mathrm{acc}}$, for details about this parameter, see Section \ref{parameters}). The core size is calculated by taking the mean \st{total} density ($\langle \rho_{\mathrm{core}}\rangle$) and mass ($M_{\mathrm{core}}$) of the core at the time measured, and approximating the core as a sphere, so $r_{\mathrm{core}} = \sqrt[3]{3M_{\mathrm{core}} /4 \uppi \langle \rho_{\mathrm{core}}\rangle}$.

We define a core as well-resolved when the ratio $r_{\mathrm{core}}/r_{\mathrm{acc}}$ is greater than or equal to 4, ensuring that $r_{\mathrm{core}}$ is sufficiently larger than $r_{\mathrm{acc}}$. The evolution of this ratio for all models and core definitions (using different threshold densities) shows that all cores are well resolved, except for the core defined with $n_{\rm thr} = 10^5$ cm$^{-3}$ in the C100 model and a few points from other cores, where $r_{\mathrm{core}}/r_{\mathrm{acc}} \gtrsim 2$. Therefore, numerical effects related to sink formation and accretion can be disregarded in nearly all cases.

\subsection{Limitations} \label{sec:limitations}

The main challenges in this work stem from the nature of the simulations. The spherical collapse occurs within a square box, which introduces border effects at late times that deviate from the idealized spherical collapse. These effects can  distort the core's geometry and impact various measurements, particularly in simulations with OBCs, where the continuous injection of material eventually leads to spurious density structures. To minimize these issues, we avoid measurements when edge effects are evident and restrict our calculations to times $t \lesssim 1.1 \tau_{\rm ff}$ for the low-density runs C100 and O100, and to times $t \lesssim 0.85 \tauff$ for the high-density runs C1000 and O1000.

On the other hand, although our open-boundary-condition simulations attempt to capture the effect of accretion from the parent clump or filament onto the core, they only do so at an approximate level, since the accretion flow is {\it dragged} by the interior collapse flow, rather than occurring as a consequence of the global self-gravity of the system. So, our open-boundary cases are to be considered as a lower limit to the effect of external accretion onto the cores.

\section{{\bf Summary and} Conclusions} \label{sec:summary}

In this work, we carried out a set of simplified simulations of the gravitational collapse of an isolated, self-gravitating core undergoing spherical collapse, varying the initial mean number density ($n_0 = 100$ and $1000$ cm$^{-3}$). These two density regimes allowed us to model both low-mass cores (models O100 and C100) and high-mass cores (models O1000 and C1000), each evolved under open or closed boundary conditions (BCs). We examined how the definition of the core (set by different density thresholds, $n_{\rm thr}$) affects its physical properties and inferred star formation activity. We also investigated the role of material replenishment by comparing simulations with open and closed BCs, which respectively allow or inhibit the accretion of external gas into the computational domain. Through these experiments, we assessed how both the core definition and BCs influence the star formation rate (SFR), the star formation efficiency (SFE), and the ratio of the SFR to the gas-infall rate ($\epsilon_{\rm ff}$). Our main findings and conclusions can be summarized as follows:

\begin{itemize}
    
    \item In all models, the sink accretion rate accelerates during the early stages of sink evolution. In the more massive cores (models O1000 and C1000), the SFR is at least one order of magnitude higher than in the low-mass cores (models O100 and C100).

    \item The SFE depends on the core definition, the adopted BCs, and the evolutionary state of the system. In the low-density models (C100 and O100), the SFE grows rapidly as the cores evolve and their gas reservoirs are depleted or replenished depending on the BCs, reaching values close to unity. In contrast, in the high-density models (C1000 and O1000), the SFE remains $\lesssim 0.8$, with larger values attained in more evolved and more compact cores defined with higher density thresholds. However, this difference is partly a consequence of the high-density runs being followed to a less advanced stage of collapse, since they had to be stopped at earlier free-fall times for numerical reasons.
    
    \item We measure $\epsilon_{\rm ff}$ using the average SFR, the free-fall time, and core masses defined at different density thresholds ($n_{\rm thr}$). As shown in Fig.~\ref{fig:SFRff_evol}, $\epsilon_{\rm ff}$ rises immediately after sink formation and then remains relatively stable while gas continues to be replenished, with values of $\sim 0.001 - 0.3$ in the high-density models (C1000 and O1000) and $\sim 0.01 - 0.5$ in O100, reflecting a comparable growth of the SFR and the core gas infall rate. In contrast, C100, which we follow to a more advanced stage of collapse, shows a continuous rise in $\epsilon_{\rm ff}$ as its core mass decreases and, without external gas replenishment, the infall rate drops. The choice of $n_{\rm thr}$ affects all models, with higher thresholds yielding higher $\epsilon_{\rm ff}$, a trend also present in the radial profiles (Fig.~\ref{fig:SFRff_PROF}). These profiles shift upward in time while maintaining their shape and generally decline slightly with radius Counterintuitively, $\epsilon_{\rm ff}$ is systematically higher in the low-mass cores, because the larger gas infall rates of the high-mass cores compensate for their higher SFRs, indicating that the gas infall rate, rather than the core mass alone, plays a dominant role in setting $\epsilon_{\rm ff}$.
    
\end{itemize}

 Our results emphasize that accretion from the surrounding medium is an important ingredient in the evolution of collapsing cores. Because the cores continuously grow in mass while forming stars, the gas reservoir is not fixed but evolves together with the stellar component. Consequently, the value of $\epsilon_{\rm ff}$ reflects the competition between star formation and gas accretion, rather than the efficiency of star formation alone. This behaviour is naturally expected in the hierarchical gravitational collapse scenario \citep[e.g.,][]{Vazquez_Semadeni+19}, where structures at all scales are fed by inflows from their larger-scale environment.

The previous findings lead to several important conclusions. The measured SFR, SFE, and $\epsilon_{\rm ff}$ all depend sensitively on how the core is defined, as different density thresholds ($n_{\rm thr}$) select regions with distinct mass reservoirs, free-fall times, and evolutionary states. Higher density thresholds yield more compact cores with higher SFE and $\epsilon_{\rm ff}$. The boundary conditions (BCs) also exert a significant influence: simulations with open BCs, which allow the injection of fresh material, tend to sustain higher SFRs but yield lower SFE and lower $\epsilon_{\rm ff}$, because the core mass grows in tandem with the sink accretion. In contrast, closed BCs prevent mass replenishment, leading to a rapid decline in core mass and, consequently, high SFE and $\epsilon_{\rm ff}$. These results highlight that both core definition and BCs shape the inferred star formation properties, and must therefore be carefully accounted for when comparing numerical models with observations.

\section*{Acknowledgements}

AP and EVS acknowledge financial support from the UNAM-PAPIIT IG100223 grant. 
JBP acknowledges financial support from the UNAM-PAPIIT grant number IN110026.
AP also acknowledges support from the SNII of SECIHTI, M\'exico. GCG acknowledges support from UNAM-PAPIIT grant number IN110824. 
The authors thankfully acknowledge computer resources, technical advice and support provided by LANCAD-UNAM-DGTIC-188 and SECIHTI, through the use of the Miztli supercomputer at DGTIC–UNAM.  Much of the data analysis and visualization presented in this work were carried out using the \texttt{yt} toolkit \citep{yt}.

\section*{Data Availability}

The data underlying this article will be shared on reasonable request to the corresponding author.



\bibliographystyle{mnras}
\bibliography{ref} 








\bsp	
\label{lastpage}
\end{document}